\documentclass[pre,twocolumn,preprintnumbers,showpacs]{revtex4}
\usepackage{graphicx}

\begin{document}

\title{Monotonic entropy growth for a nonlinear model of random exchanges}

\author{S. M. Apenko }
\affiliation{ I E Tamm Theory Department, P N Lebedev Physical Institute, Moscow,
119991, Russia} \email{apenko@lpi.ru}

\begin{abstract}
We present a proof of the monotonic entropy growth for a nonlinear discrete-time model
of a random market. This model, based on binary collisions, also may be viewed as a
particular case of Ulam's redistribution of energy problem. We represent each step of
this dynamics as a combination of two processes. The first one is a linear
energy-conserving evolution of the two-particle distribution, for which the entropy
growth can be easily verified. The original nonlinear process is actually a result of
a specific `coarse-graining' of this linear evolution, when after the collision one
variable is integrated away. This coarse-graining is of the same type as the real
space renormalization group transformation and leads to an additional entropy growth.
The combination of these two factors produces the required result which is obtained
only by means of information theory inequalities.
\end{abstract}


\maketitle

It is widely known that for a stochastic Markov process described by a linear master
equation for a distribution function $p(x,t)$ defined on some space of $x$'s there
exists a Lyapunov function which monotonically decreases while we approach
equilibrium. This function is just the relative entropy $K=\sum_{x}p(x,t)\ln
p(x,t)/p_0(x)$, where $p_0(x)$ is an equilibrium distribution \cite{Sch} (see also
earlier works by Schl\"{o}gl \cite{Schl} and, e.g., \cite{rec} for more recent related
studies). This relative entropy has the meaning of the `information gain' that we
obtain when the knowledge about the true distribution $p(x,t)$ becomes available if we
{\em a priory} knew only the equilibrium one $p_0(x)$. $K$ may be also viewed as a
total entropy production that takes place during the whole relaxation process
\cite{Sch,Schl,rec}.

There is no such general result for nonlinear evolution equations, where monotonicity
of possible Lyapunov functions should be proved independently for each problem. There
are, however, situations when it is just the Boltzmann entropy which monotonically
grows when we approach equilibrium, the most known example being the Boltzmann
equation itself where this monotonicity is established by the famous $H$-theorem
\cite{BE}.

Quite recently an interesting nonlinear evolution was proposed and analyzed as a
gas-like economic model in a series of papers \cite{gl}. This is a discrete-time
evolution for distributions $p(x)$ with continuous $x\geq 0$ and on each step of
iterative procedure $p(x)\rightarrow p'(x)$, where
\begin{equation}\label{gl}
  p'(x)=\int_0^{\infty}\int_0^{\infty}dudv \,\frac{\theta(u+v-x)}{u+v}p(u)p(v)
\end{equation}
and the $\theta$-function ensures that $u+v>x$. This model assumes that economic
transactions occur by binary `collisions' between agents who exchange money in the
same way as particles in a gas exchange their energy \cite{gas} and after each
collision the total amount of money they both possessed is distributed between them
absolutely at random. For initial distributions with finite mean `energy' $\langle
x\rangle$ this process converges to exponential equilibrium distribution
$p_0=\alpha\exp(-\alpha x)$, where $1/\alpha=\langle x\rangle$ \cite{gl}.

The structure of Eq. (\ref{gl}) is very transparent: We first randomly choose two
values $u$ and $v$ with probability $p(u)p(v)$, then multiply it by a transition
probability $W(u,v\rightarrow x)$, given here by $W(x)=1/(u+v)$ for $0<x<u+v$, and
finally sum over all possible choices of $u,v$ to obtain the new distribution $p(x)$.
Thus for uniform probability density $W(x)$ the factor $1/(u+v)$ in Eq. (\ref{gl})
arises simply from the normalization condition $\int W(x)dx=1$. It is easy to prove
also that $\langle x\rangle$ is conserved under this nonlinear transformation
\cite{gl}.

In fact, this process is an example of what is known as {\em Ulam's redistribution of
energy problem}, stated as follows: "Consider a vast number of particles and let us
redistribute the energy of these particles... First, pair the particles at random.
Second, for each pair, redistribute the total energy of the pair between these
particles according to some given fixed probability law of redistribution..."
\cite{U}. Ulam believed that the distribution of energy would then converge to some
final distribution independent of the initial one and later his conjecture was indeed
proved in \cite{U}. For uniform redistribution law this process is essentially the
same as the money exchanges described by Eq. (\ref{gl}). However, the nonlinear
transformation (\ref{gl}) first introduced in \cite{gl} in an economic context seems
more suited for our study than the equation for the moments of $p(x)$ used in
\cite{U}.

Since the process (\ref{gl}) is very similar in spirit to the evolution that leads to
the Boltzmann equation we expect that the entropy $S(p)=-\int dxp(x)\ln p(x)$ should
monotonically increase under this transformation. While this conjecture was first
formulated already in \cite{gl}, the analytical proof of this growth seems to be still
lacking, mainly because standard methods do not directly work for discrete-time
evolution. Note, that since entropy is obviously maximized by the exponential
distribution $p_0\sim\exp(-\alpha x)$ under the constraint $\langle
x\rangle=\mathrm{const}$ (see e.g. \cite{J}) it is monotonicity that has to be proved.

It should be noted here that for non-uniform redistribution laws in Ulam's problem we
do not expect that entropy always grows. Indeed, in the general case the limiting
distribution is no longer exponential \cite{U}, hence the entropy is not maximal in
equilibrium. A simple example is a special law when the total energy of colliding
particles is shared equally among them. In this case all particles will have the same
energy in equilibrium \cite{U} and the entropy definitely gets lower during
relaxation. For this reason here we consider only uniform redistribution described by
random market model of Eq. (\ref{gl}).

One can, of course, try to rewrite Eq. (\ref{gl}) in a form similar to Markov chain
evolution
\begin{equation}\label{mar}
  p'(x)=\int_0^{\infty}du\,\mathrm{P}(x,u;p)p(u),
\end{equation}
where `transition probability'
\begin{equation}\label{tran}
  \mathrm{P}(x,u;p)=\int_0^{\infty} dv\,\frac{\theta(u+v-x)}{u+v}p(v)
\end{equation}
itself depends on $p(x)$. Stochastic processes that may be related to such equations
are now sometimes called {\em nonlinear Markov processes} \cite{ND} though this
terminology was criticized in \cite{McC}. Regardless of what we call it, if we
substitute some solution $p(x)$ of Eq. (\ref{gl}) into $\mathrm{P}(x,u;p)$ we will end
with the linear equation (\ref{mar}) but with time dependent transition probabilities.
Close to equilibrium we may take $\mathrm{P}\simeq \mathrm{P}(x,u;p_0)$ which now
satisfies detailed balance condition
$\mathrm{P}(x,u;p_0)/\mathrm{P}(u,x;p_0)\sim\exp(-\alpha x+\alpha u)$ and hence will
definitely lead to the monotonic entropy growth. But far from equilibrium this
approach seems to be of little help.

For this reason in this note we will give a proof of the monotonic entropy growth for
Eq. (\ref{gl}) within quite a different approach, which is based almost entirely on
known information theory inequalities and utilizes the fact that certain
coarse-graining transformations always result in the entropy growth.

The main idea of the proof is (i) to introduce an auxiliary {\em linear} evolution,
defined on a larger space of two variables, for which entropy growth can be easily
proved and then (ii) to show that (\ref{gl}) is actually a result of a certain {\em
coarse-graining} of this linear evolution.

For this purpose let us introduce a `two-particle' distribution function $f(x,y)$
which after one step of evolution transforms into $f'(x,y)$,
\begin{equation}\label{lin}
  f'(x,y)=\int_0^1 d\xi\,f(\xi(x+y),(1-\xi)(x+y)).
\end{equation}
This is obviously a linear transformation and it is easy to see that it conserves
positivity of $f(x,y)$, its norm  and the mean `energy' $\langle x+y\rangle$.

The physical meaning of Eq. (\ref{lin}) is rather clear since $f(x,y)$ describes pairs
of particles. Particles, which after the collision have energies $x$ and $y$, before
the collision might have any energies $u$ and $v$ provided $u+v=x+y$ i.e. we may take
$u=\xi(x+y)$ and $v=(1-\xi)(x+y)$, where $0<\xi<1$ denotes a fraction of the total
energy that the first particle had. Then Eq. (\ref{lin}) is just the sum over all
possibilities (all of them having equal probabilities) that result in the values $x$
and $y$.

It should be noted, however, that (\ref{lin}) alone does not describe correctly
evolution of the two-particle probability distribution in Ulam's problem. It takes
into account only collisions within fixed pairs of particles and if we choose initial
distribution as a $\delta$-function localized at some point $(x_0$, $y_0)$ then after
the first iteration it will be uniformly smeared along the isoenergetic line
$x+y=x_0+y_0$ and will not change afterwards. Though the expected true two-particle
equilibrium distribution
\begin{equation}\label{equ}
f_0(x,y)=\alpha^2\exp[-\alpha(x+y)]
\end{equation}
is certainly a fixed point of the transformation (\ref{lin}), this evolution alone
cannot explain relaxation to (\ref{equ}) from arbitrary initial $f(x,y)$ because this
requires new random pairings of particles at each step, not included in (\ref{lin}).
That is why (\ref{lin}) has a lot of additional spurious `equilibriums'---any function
that depends only on $x+y$ does not change under this transformation. For these
reasons only one iteration of Eq. (\ref{lin}) really makes sense and its only purpose
is to produce nonlinear equation (\ref{gl}) after some `projection' procedure,
described below.

But let us first show that entropy grows for the transformation (\ref{lin}). The
regular way to prove such monotonicity theorems is to start from the relative entropy
or Kullback-Leibler (KL) distance $D(\mu||\nu)=\sum\mu\ln\mu/\nu$ between two
probability distributions $\mu$ and $\nu$. It is well known from information theory
that $D(\mu||\nu)$ cannot increase under `coarse-graining' of these distributions,
when some variables are integrated out. This immediately follows from the chain rule
for relative entropy \cite{inf} and some examples of how this works may be found e.g.
in \cite{KL,a}. Consider now the distribution
\begin{equation}\label{mu}
  \mu(\xi,x,y)=f(\xi(x+y),(1-\xi)(x+y)),
\end{equation}
defined on the space $\xi\in[0,1]$, $x,y\in[0,\infty)$ and define $\nu(\xi,x,y)$ in
the same way through the equilibrium distribution $f_0(x,y)$ from (\ref{equ}). It is
easy to check that both $\mu$ and $\nu$ are positive and normalized to unity. Then
define the coarse-graining procedure $\mu\rightarrow\tilde{\mu}$ as averaging over the
$\xi$ variable, i.e.
\begin{equation}\label{mucg}
\tilde{\mu}(x,y)=\int_0^1 d\xi\,\mu(\xi,x,y)=f'(x,y),
\end{equation}
according to Eq. (\ref{lin}), and, obviously, $\tilde{\nu}(x,y)=f_0(x,y)$.

The above statement about the monotonic behavior of KL distance can be written as
\begin{eqnarray}\label{KL}
  \int_0^1 d\xi\int_0^{\infty} dxdy\,\mu(\xi,x,y)\ln\frac{\mu(\xi,x,y)}{\nu(\xi,x,y)}\geq
  \nonumber \\
  \geq\int_0^{\infty} dxdy\,\tilde{\mu}(x,y)\ln\frac{\tilde{\mu}(x,y)}{\tilde{\nu}(x,y)}
\end{eqnarray}
In the integral on the left-hand side we now make a change of variables
\begin{equation}\label{cv}
  u=\xi(x+y),\quad
v=(1-\xi)(x+y),\quad z=x-y
\end{equation}
with the obvious property $x+y=u+v$. Then the integration measure and ranges of
integration transform as follows
\begin{equation}\label{int}
  \int_0^1 d\xi\int_0^{\infty}dxdy=
  \frac{1}{2}\int_0^{\infty}dudv\int_{-(u+v)}^{(u+v)}dz\frac{1}{u+v}
\end{equation}
and since according to (\ref{mu}) the integrand does not depend on $z$, integration
over $z$ exactly cancels the Jacobian $1/2(u+v)$.

Then, using also exact expressions for $\tilde{\mu}$, Eq. (\ref{mucg}), and
$\tilde{\nu}$, we can rewrite Eq. (\ref{KL}) as
\begin{eqnarray}\label{mon}
  \int_0^{\infty} dudv\, f(u,v)\ln\frac{f(u,v)}{f_0(u,v)}\geq \nonumber\\
  \geq\int_0^{\infty} dxdy\, f'(x,y)\ln\frac{f'(x,y)}{f_0(x,y)}
\end{eqnarray}
Thus the relative entropy could not increase under the transformation (\ref{lin}).
Certainly the same inequality will be valid if we substitute any normalized fixed
point solution of (\ref{lin}) instead of the exponential distribution $f_0(x,y)$, but
choosing $f_0$ is more suitable for what follows.

The same way of reasoning may be applied actually for any linear Markov evolution of
the form (\ref{mar}) with $\mathrm{P}(x,u)$ independent of $p(x)$. One should take
$\mu(x,u)=\mathrm{P}(x,u)p(u)$, $\nu(x,u)=\mathrm{P}(x,u)p_0(u)$, where $p_0(u)$ is an
equilibrium distribution, and then integrate  them over $u$ to obtain
$\tilde{\mu}=p'(x)$ and $\tilde{\nu}=p_0(x)$. Then from equation similar to Eq.
(\ref{KL}) it follows that $K=\int dxp(x)\ln p(x)/p_0(x)$ monotonically decreases on
each iteration, which is, of course, well known. The main idea behind this proof is
that any such evolution may be viewed as some kind of coarse-graining since it
includes integration over initial data (cf. \cite{ap}). Note that for nonlinear
evolution, when $\mathrm{P}(x,u;p)$ depends on the distribution function $p(u)$, as in
(\ref{mar}), and hence changes on each step, this approach does not work, because now
$p_0$ is not an instantaneous equilibrium and $\tilde{\nu}\neq p_0(x)$.

Now, since we have chosen the equilibrium distribution in the exponential form,
$f_0\sim\exp[-\alpha(x+y)]$, Eq. (\ref{mon}) may be rewritten, as usual, as $F\geq
F'$, where $F=\langle x+y\rangle-S/\alpha$ is the free energy and the entropy is given
by $S(f)=-\int dxdyf(x,y)\ln f(x,y)$. But our linear transformation conserves the mean
energy $\langle x+y\rangle$, hence the monotonicity of the free energy results in the
entropy growth
\begin{equation}\label{S}
  S'\geq S.
\end{equation}
Thus we have proved the monotonic entropy growth for our auxiliary linear
transformation (\ref{lin}). This is an almost evident result and it is only the first
step of the proof for the original nonlinear problem.

Now we need to relate the linear process (\ref{lin}) to the initial nonlinear
evolution (\ref{gl}). For this purpose consider in Eq. (\ref{lin}) a special
factorized initial condition
\begin{equation}\label{init}
 f(x,y)=p(x)p(y)
\end{equation}
and define the transformed probability $p'(x)$ as the marginal probability for the
transformed distribution, i.e.
\begin{equation}\label{prime}
  p'(x)\equiv\int_0^{\infty} dy f'(x,y).
\end{equation}
Since $f'(x,y)$ is symmetric under the permutation of variables $x$ and $y$ it
actually does not matter which one variable to integrate out.

Then we have
\begin{eqnarray}\label{nonl}
  p'(x)=\int_0^{\infty}dy\int_0^1 d\xi\, p(\xi(x+y))p((1-\xi)(x+y))= \nonumber\\
  =\int_0^{\infty}du\int_0^{\infty}dv\,\frac{\theta(u+v-x)}{u+v}p(u)p(v),\qquad
\end{eqnarray}
where we have made the change of variables $(\xi,y)\rightarrow (u,v)$ similar to
(\ref{cv}), $u=\xi(x+y)$, $v=(1-\xi)(x+y)$, with the Jacobian $1/(u+v)$. The condition
$u+v>x$ arises from the positivity of $y=u+v-x$. Clearly this is exactly the required
nonlinear Eq. (\ref{gl}).

Thus, on each step the nonlinear evolution of the gas-like model may be obtained by a
kind of `projection' procedure from the linear transformation (\ref{lin}) by choosing
the special initial conditions (\ref{init}) and by the subsequent elimination of one
variable from the resulting two-particle distribution function (\ref{prime}).

Another way to look at this phenomenon is to say that our non-linear evolution may be
represented as a combination of two processes. The first one is the linear evolution
of Eq. (\ref{lin}) with initial condition (\ref{init}) which should be supplemented
then by the subsequent `reduction' of $f'(x,y)$ back to the factorized form, which
corresponds to a new random pairing of particles and is, in its turn, the new initial
condition for the next step.

But elimination of exactly half of the variables, as in Eq. (\ref{prime}), is also
related to some monotonicity property. For example, for the real space decimation
renormalization transformation in spin systems, when on each step of renormalization
we divide the lattice into two identical sublattices and half of all spins are summed
away \cite{renorm}, the entropy per lattice site was shown to grow monotonically
\cite{a}. For the sake of completeness we repeat here this simple derivation as
applied to our present system.

This monotonicity of entropy per degree of freedom results just from the positivity of
the mutual information of two sets of variables. In our present case the mutual
information of $x$ and $y$ variables after the transformation (\ref{lin}), whose joint
probability distribution is $f'(x,y)$ and marginal distributions are $p'(x)$ and
$p'(y)$, is given by the usual formula \cite{inf}
\begin{equation}\label{mut}
  I=\int\int dxdy\,f'(x,y)\ln\frac{f'(x,y)}{p'(x)p'(y)}\geq 0.
\end{equation}
This mutual information may be written also as a difference between the sum of
entropies of subsystems (which are identical in our case) and the total entropy of the
joint distribution
\begin{equation}\label{mutu}
  I=2S(p')-S(f'),
\end{equation}
where $S(p')=-\int dxp'(x)\ln p'(x)$ and $S(f')$ is the entropy of the two particle
system which earlier in Eq. (\ref{S}) was denoted by $S'$.

Hence from $I\geq 0$ it follows
\begin{equation}\label{SS}
  S(p')\geq \frac{1}{2}S'.
\end{equation}
Note that this inequality is not just a trivial consequence of the information loss or
decrease of the relative entropy after one variable is eliminated. Information loss
results in the {\em decrease of the total entropy}, which looks like $S'\geq S(p')$
\cite{a} and is clearly distinct from Eq. (\ref{SS}).

Now we can combine this inequality with the one obtained earlier for the linear
evolution, Eq. (\ref{S}), to arrive at $S(p')\geq S/2$. But for the factorized initial
distribution Eq. (\ref{init}) the entropy $S$ is just twice the one-particle entropy
of the distribution $p(x)$, i.e. $S=2S(p)$ and hence we finally have
\begin{equation}\label{fin}
  S(p')\geq S(p).
\end{equation}
This completes the proof that the entropy $S(p)=-\int dxp(x)\ln p(x)$ monotonically
grows on each step under iterations of the nonlinear transformation (\ref{gl}).

Let us now give an example illustrating our general proof. If we start from the
distribution $p(x)=x\exp(-x)$ with entropy $S(p)=\gamma+1\simeq 1.5772$ ($\gamma\simeq
0.5772$ is the Euler constant), then for the factorized initial condition (\ref{init})
we have $f'(x,y)=1/6(x+y)\exp[-(x+y)]$ from Eq. (\ref{lin}). The corresponding entropy
per degree of freedom is now larger, $S'/2=\gamma+1/6+1/2\ln(6)\simeq 1.6397>S(p)$.
After we eliminate one variable we finally have $p'(x)=1/6(x^2+2x+2)\exp(-x)$ (see
also \cite{gl}) and the entropy now equals $S(p')\simeq 1.6667$ which in its turn is
slightly larger than $S'/2$. Thus we see how entropy indeed grows on each stage of our
combined evolution that is equivalent to the initial nonlinear transformation.

In summary, we have proved the monotonic entropy growth for a nonlinear evolution
which describes pairwise interaction of economical agents with random money exchanges
and also may be viewed as a particular case of Ulam's redistribution of energy
problem. The proof is based on representing a single step of the nonlinear evolution
as a combination of two steps: The first is related to an auxiliary linear
two-particle process and the second one is a kind of a coarse-graining, similar to
decimation renormalization transformation, when one of the two variables is integrated
away. Since on both steps the entropy can be shown to increase we conclude that the
entropy is indeed monotonically increasing for the original nonlinear problem.

The proof is based entirely on information theory inequalities and possibly may be of
some use for other nonlinear problems. It is not clear however whether it is possible
to use the present approach or some of its modifications to find Lyapunov functions
for non-uniform redistribution laws in the general Ulam problem.

I am very grateful to J. Gaite for pointing out Ref. \cite{U} and valuable comments,
to R. L\'{o}pez-Ruiz for stimulating correspondence and important remarks, to V.
Losyakov for many discussions and to J.L. McCauley for sending me his papers. The work
was supported in part by RFBR Grants No. 10-02-00509, 11-02-90453 and 12-02-00520.

\end{document}